\documentclass[aps,prl,twocolumn,superscriptaddress,10pt,article,showpacs,longbibliography]{revtex4-2}

\usepackage{xcolor}
\usepackage{graphicx}
\usepackage{dcolumn}
\usepackage{bm}

\usepackage[normalem]{ulem}
\usepackage{float}
\usepackage{multirow}
\newcommand{\vbm}[0]{$\mathrm{V}_{\mathrm{B}}^-$}
\newcommand{\nfour}[0]{${}^{14}\mathrm{N}$}
\newcommand{\nfive}[0]{${}^{15}\mathrm{N}$}

\begin{document}

\preprint{APS/123-QED}

\title{Intrinsic high-fidelity spin polarization of charged vacancies in hexagonal boron nitride}

\author{W. Lee}
\thanks{These authors contributed equally to this work.}
\affiliation{Department of Physics, Harvard University, Cambridge, Massachusetts 02138, USA}

\author{V. S. Liu}
\thanks{These authors contributed equally to this work.}
\affiliation{Department of Physics, Harvard University, Cambridge, Massachusetts 02138, USA}

\author{Z. Zhang}
\thanks{These authors contributed equally to this work.}
\affiliation{Department of Physics, Harvard University, Cambridge, Massachusetts 02138, USA}

\author{S. Kim}
\affiliation{Department of Physics, Harvard University, Cambridge, Massachusetts 02138, USA}

\author{R. Gong}
\affiliation{Department of Physics, Washington University, St. Louis, Missouri 63130, USA}

\author{X. Du}
\affiliation{Department of Physics, Washington University, St. Louis, Missouri 63130, USA}

\author{K. Pham}
\affiliation{Department of Physics, Harvard University, Cambridge, Massachusetts 02138, USA}

\author{T. Poirier}
\affiliation{Tim Taylor Department of Chemical Engineering, Kansas State University, Manhattan, Kansas 66506, USA\looseness=-1}

\author{Z. Hao}
\affiliation{Department of Physics, Harvard University, Cambridge, Massachusetts 02138, USA}

\author{J. H. Edgar}
\affiliation{Tim Taylor Department of Chemical Engineering, Kansas State University, Manhattan, Kansas 66506, USA\looseness=-1}

\author{P. Kim}
\affiliation{Department of Physics, Harvard University, Cambridge, Massachusetts 02138, USA}

\author{C. Zu}
\affiliation{Department of Physics, Washington University, St. Louis, Missouri 63130, USA}

\author{E. J. Davis}
\affiliation{Department of Physics, Harvard University, Cambridge, Massachusetts 02138, USA}
\affiliation{Department of Physics, New York University, New York, New York 10003, USA}

\author{N. Y. Yao}
\affiliation{Department of Physics, Harvard University, Cambridge, Massachusetts 02138, USA}

\date{\today}

\begin{abstract}

The negatively charged boron vacancy (\vbm) in hexagonal boron nitride (hBN) has garnered significant attention among defects in two-dimensional materials.
This owes, in part, to its deterministic generation, well-characterized atomic structure, and optical polarizability at room temperature.
We investigate the latter through extensive measurements probing both the ground and excited state polarization dynamics.
We develop a semiclassical model based on these measurements that predicts a near-unity degree of spin polarization, surpassing other solid-state spin defects under ambient conditions.
Building upon our model, we include the presence of nuclear spin degrees of freedom adjacent to the \vbm~and perform a comprehensive set of Lindbladian numerics to investigate the hyperfine-induced polarization of the nuclear spins. 
Our simulations predict a number of important features that emerge as a function of magnetic field which are borne out by experiment.

\end{abstract}

\maketitle


In the landscape of existing quantum platforms, localized defects in solids are controlled quantum degrees of freedom that offer unique opportunities at the atomic scale~\cite{weber2010, atature2018, awschalom2018, wolfowicz2021}. 
These include the sensing of strain and magnetism in materials~
\cite{balasubramanian2008, doherty2014,  kraus2014}, strong light-matter coupling in photonic cavities for quantum networks~\cite{togan2010, sipahigil2016, lukin2020, ruskuc2022}, and the quantum simulation of disordered spin systems~\cite{choi2017, randall2021, davis2023}. There are many host materials for such defects, ranging from diamond and silicon-carbide to gallium nitride and an assortment of rare-earth-doped materials~\cite{doherty2013, koehl2011, nagy2019, zhong2019, luo2024}; distinct hosts provide distinct advantages including large band-gaps, strong spin-orbit coupling, or direct integration with existing nanofabrication toolsets~\cite{atature2018, awschalom2018, wolfowicz2021}.

The last several years have seen a tremendous amount of attention  focused on optically-active defects in two-dimensional van der Waals materials~\cite{liu2019, liang2021, azzam2021, tetienne2021}.
This owes, in part, to the observation of a remarkably diverse array of strongly-correlated physical phenomena in such systems~\cite{bolotin2009,cao2018mott, cao2018superconductor, wang2020, regan2020}. 
A bit more intrinsically, as defect hosts, two-dimensional materials also exhibit a number of favorable features, including: (i) ease of integration into electronic and photonic devices, (ii) well-developed fabrication techniques and (iii) in the context of quantum sensing, the ability to provide sub-nanometer proximity to the sample being probed~\cite{liu2019, azzam2021, tetienne2021}.

While initial efforts on defects in 2D materials centered around developing tunable single-photon emitters~\cite{srivastava2015, koperski2015, chakraborty2015, he2015, tran2016}, seminal recent works have identified a number of spinful defects exhibiting optically 
detectable magnetic resonance (ODMR)~\cite{gottscholl2020,  stern2022, stern2024}. Perhaps the most well-known of these is the negatively charged boron vacancy (\vbm) in hexagonal boron nitride (hBN).
Being a spin-1 defect optically-polarizable at room temperature, the \vbm~exhibits a number of features reminiscent of the paradigmatic nitrogen-vacancy center in diamond~\cite{gottscholl2020,abdi2018,jelezko2002,doherty2013}.
That hBN is commonly used to
\begin{figure}[H]
    \centering
\includegraphics[width=1.0\columnwidth]{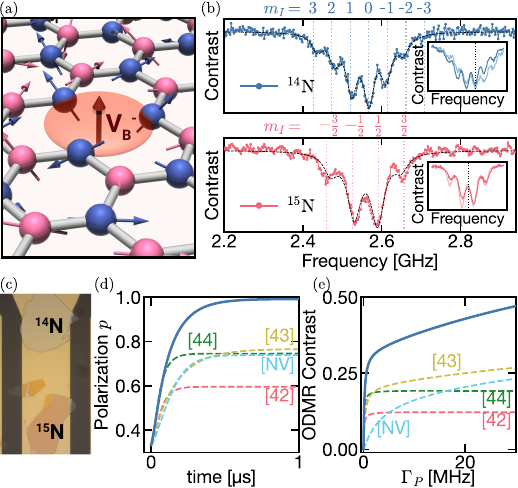}
    \caption{(a) Schematic depicting the structure of the \vbm~defect in hBN. (b) Continuous-wave ODMR spectra at $B_z = 32~$mT for hBN samples with naturally-occurring \nfour~(top) and isotopically-purified \nfive~(bottom)~\cite{clua2023,gong2024}. Inset: ODMR near ESLAC ($B_z = 74~$mT) at laser power $P = 4~$mW (lighter lines) and $P = 45~$mW (darker lines). At high powers, the relative amplitudes of the ODMR peaks shift owing to the nuclear spins becoming partially polarized.
    (c) Microscope image of the natural (\nfour) hBN and isotopically-purified (\nfive) hBN flakes transferred onto a coplanar waveguide~\cite{supp}. (d) Predicted time evolution of the electronic spin polarization under optical pumping at rate, $\Gamma_P = 20~$MHz~\cite{supp}.
    The solid blue line represents the result obtained with the rates shown in Table~\ref{tab:table1}, using $r = 0$. 
    For comparison, other dashed lines are plotted based on the estimated rates from \cite{baber, whitefield, jacques}; the dashed light-blue line represents the case of the nitrogen-vacancy center~\cite{Zu2021,Robledo_2011,Tetienne_2012}. (e) ODMR contrasts as a function of the optical pumping rate $\Gamma_P$. Here, we take $\Gamma_P^\mathrm{NV-}/\Gamma_P^{\mathrm{V}_{\mathrm{B}}^-} \sim 0.1$ \cite{supp}.}
    \label{fig1}
\end{figure}
\noindent encapsulate van der Waals heterostructures, has led to an immediate application for \vbm~as an ideal, local sensor of magnetism in quantum materials~\cite{kumar2022, huang2022, zhou2024, healey2023}.

Despite these applications and intense interest in the metrological capabilities of \vbm~\cite{kumar2022, huang2022, healey2023, zhou2024, gottscholl2021sensing, lyu2022, liu2021, udvarhelyi2023, Durand2023}, many fundamental properties about the defect remain unknown. 
Indeed, while experimental and theoretical studies have made marked progress~\cite{reimers2020, ivady2020, baber, whitefield, jacques, tongcang, mathur2022, gottscholl2021odmr, gao2022, murzakhanov2022, lee2022, haykal2022, mu2022, gong2023, Durand2023}, a detailed understanding of the polarization and charge dynamics, intersystem crossing (ISC) rates, singlet lifetimes, and nuclear spin dynamics, is still outstanding. 
This lack of a full microscopic understanding leads to a number of broader challenges: first, are there intrinsic properties of \vbm~that stand out relative to other ambiently-polarizable spin defects? 
Second, what specific features of \vbm~must be improved to enhance its performance as a quantum sensor, or to enable the detection of single defects?

In this Letter, we take an important step toward answering these questions by combining an extensive set of ground and excited state spin dynamics measurements (for both \nfour~and \nfive~defect isotopes) with detailed Linbladian simulations (Fig.~\ref{fig1})~\cite{baber, whitefield, jacques}.
Our main results are twofold. 
First, we develop a semiclassical model for the photo-induced spin dynamics of \vbm~ensembles~\cite{baber, whitefield, jacques}. 
Our measurements allow us to refine a number of key parameters in such models, including, for example, the inter-system crossing rate from the excited $|m_s = 0\rangle$ state relative to the $|m_s = \pm 1 \rangle$ states. 
Although our extracted rates are in tension with \emph{ab-initio} predictions~\cite{ivady2020,reimers2020}, they immediately help to resolve an open question regarding the origin of the very high ODMR contrasts that have been observed [Fig.~\ref{fig1}(e)]~\cite{tongcang}. 
Moreover, we also conjecture that  \vbm~exhibits at least one stand-out property: namely, that off-resonant excitation under ambient conditions leads to an extremely high degree of spin polarization with $p \gtrsim 95\%$ [Fig.~\ref{fig1}(d)]. 
Such high polarizations could be especially important for quantum simulation using  \vbm~ensembles~\cite{randall2021, wang2015}.

Our second main result concerns the polarization dynamics of nuclear spins adjacent to the \vbm.
Understanding such dynamics is crucial to both: (i) mitigating the decoherence effects of the nuclear spin bath~\cite{gottscholl2021odmr, murzakhanov2022, lee2022, haykal2022, gong2023} and (ii) harnessing adjacent nuclear spins as controllable degrees of freedom~\cite{gao2022}. 
To this end, we investigate the  spin polarization timescale of \vbm~as a function of magnetic fields, with a focus on the ground- (GSLAC) and excited-state level anti-crossings (ESLAC), where hyperfine interactions significantly affect the polarization dynamics ~\cite{baber, mu2022, gao2022, smeltzer2009}. 
By incorporating nuclear spins into our semiclassical model, we are able to utilize a full Linbladian simulation to predict  essential features of such dynamics, including an isotope-dependent inversion near GSLAC. 
These predictions are indeed borne out by our data. 

\begin{figure}
    \centering
    \includegraphics[width=\columnwidth]{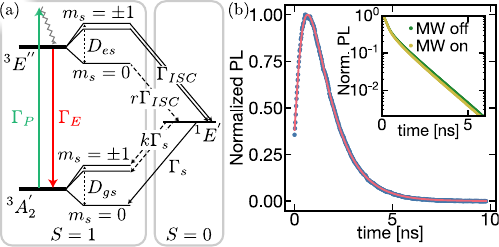}
    \caption{(a) Schematic depicting the relevant states of the  seven-level rate model for the \vbm~electronic spin dynamics. 
    (b) Differential signal of the time-resolved photoluminescence  decay. The solid line corresponds to a bi-exponential fit obtained from the rate model. 
    Inset: The time-resolved photoluminescence decay with and without a $\pi$-pulse (for additional details, see supplementary material~\cite{supp}).}
    \label{fig2}
\end{figure}

\emph{Spin polarization dynamics of \vbm~ensembles}---The \vbm~defect exhibits a spin, $S=1$, ground state ($|m_s=0,\pm1 \rangle$), which can be initialized and read out via optical excitation [Fig.~\ref{fig2}(a)].
This optical polarization cycle is central to both its properties and capabilities, and is typically driven (at ambient conditions) by off-resonant illumination using 532~nm laser light. 
The illumination is understood to drive population dynamics between three manifolds of states~\cite{gottscholl2020, ivady2020, reimers2020}: the spin-triplet ground state (${}^3 A_2^{'}$), the spin-triplet excited state (${}^3 E^{''}$), and the metastable spin-singlet  state (${}^1 E^{'}$).
In the absence of external perturbations, the $|m_s = \pm 1\rangle$ states are degenerate and split from the $|m_s = 0\rangle$ state by $D_{\mathrm{gs}} = 2\pi\times 3.48~$GHz and $D_{\mathrm{es}} = 2\pi\times2.1~$GHz, for the ground and excited states, respectively~\cite{gottscholl2020, mu2022}.

After spin-conserving excitation to the excited state (at a rate, $\Gamma_P = \alpha P$, proportional to the laser power), there are two decay channels---a spin-conserving decay, $\Gamma_E$, that directly couples the ground and excited states, and an intersystem crossing, $\Gamma_{\mathrm{ISC}}$, from the excited state to the metastable singlet.
A differential branching ratio for the latter, between $|m_s = 0\rangle$ ($r\Gamma_{\mathrm{ISC}}$) and $|m_s = \pm 1\rangle$ ($\Gamma_{\mathrm{ISC}}$), enables spin polarization to accumulate in the ground state. 
The precise amount of polarization is also controlled by a second branching ratio, from the singlet to the ground state, with decay to the $|m_s = 0\rangle$ state ($\Gamma_s$) expected to be greater than decay to the $|m_s = \pm 1\rangle$ states ($k\Gamma_s$).

Characterizing these various rates has been the subject of both experimental efforts and ab-initio theoretical modeling (Table~\ref{tab:table1}); however, it is challenging to fully constrain all of the relevant parameters with any single set of experiments, and simulating the underlying spin-orbit coupling responsible for the intersystem crossing's branching ratios is also difficult~\cite{baber, whitefield, jacques, ivady2020, reimers2020}. 
Coupled with sample-dependent background variations, this has led to uncertainty in these electronic transition rates, and more broadly in the polarization pathway of the \vbm~defect. 
This uncertainty is also highlighted by an inability to fully account for certain experimental results such as high ODMR contrast ($\gtrsim 46\%$)~\cite{tongcang} and low photoluminescence~\cite{gottscholl2020}. 

To address these questions, we combine three experimental measurements: (i) spin-dependent excited-state spectroscopy, (ii)  power-dependent photoluminescence time traces from an unpolarized initial state, and (iii) spin-resolved polarization dynamics.
The choice of these measurements is specifically designed to provide overlapping constraints on the various transition rates.
For example, the excited-state spectroscopy most naturally constrains a combination of the decay rates out of the ${}^3 E^{''}$ manifold including $\{ \Gamma_E, \Gamma_\mathrm{ISC}, r \}$, while the photoluminescence provides information on the polarization timescales controlled by $\{ \Gamma_P, \Gamma_s\}$.
Finally, the spin-resolved polarization dynamics intuitively shed light on the spin-dependent branching ratios $\{r ,k \}$, but more generally, yield global constraints on all of the transition rates. 
In all cases, we hasten to emphasize the importance of carefully accounting for the photoluminescence, e.g. from other emitters in hBN and purported charge dynamics. 

\begin{table}
\begin{center}
\begin{tabular}{ |c||c|c|c|c|c| }
  \hline
  & $\Gamma_\mathrm{ISC}$ & $\Gamma_{E}$ & $\Gamma_{s}$ & $r$ & $k$ \\
  \hline
  \hline
  This work & $1286\,\text{MHz}$ & $849\,\text{MHz}$ & $22.3\,\text{MHz}$ & $\lesssim 0.04$ & $0.21$ \\
  \hline
  \cite{whitefield}& $1150\,\text{MHz}$ & $880 \,\text{MHz}$ & $20 \,\text{MHz}$ & 0.11 & 0.65 \\
  \hline
  \cite{jacques} & $1800\,\text{MHz}$ & \multirow{3}{*}{\begin{tabular}[c]{@{}c@{}}$1/11 \,\text{MHz}$\\ (DFT)\end{tabular}} & $41 \,\text{MHz}$ & 0.46 & 0.17 \\
  \cline{1-2} \cline{4-6} 
  \cite{baber} & $2030\,\text{MHz}$ & & $20 \,\text{MHz}$ & 0.50 & 0.34 \\
  \cline{1-2} \cline{4-6} 
  \cite{reimers2020}& $588\,\text{MHz}$ & & $0.56 \,\text{Hz}$ &  &  \\
  \hline
\end{tabular}

\end{center}

\caption{\label{tab:table1}%
Extracted electronic spin transition rates for the seven-level rate model in Fig.~\ref{fig2}(a).}
\end{table}

\emph{Determining electronic transition rates}---We begin by considering the spin-dependent decay rates from the excited state~\cite{baber, jacques}.
In particular, after initializing to $|m_s=0\rangle$, we apply a $5$~ps laser pulse to drive the population to the excited state; we repeat the same measurement starting from the $|m_s=-1\rangle$ state (resolved from the $|m_s=+1\rangle$ state using a magnetic field $B_z = 12~$mT).
In each case, we measured the time-resolved decay of the photoluminescence as depicted in the inset of Fig.~\ref{fig2}(b).
Crucially, the differential signal between these measurements is independent of factors such as pulse fidelities and background photoluminescence, and thus, characteristic of only the intrinsic dynamics of \vbm.
The semiclassical model [Fig.~\ref{fig2}(a)] predicts that our differential signal should exhibit a simple bi-exponential form 
$\sim e^{-(\Gamma_E + r\Gamma_{\mathrm{ISC}})t} - e^{-(\Gamma_E + \Gamma_{\mathrm{ISC}})t}$, which indeed captures the data extremely well [Fig.~\ref{fig2}(b)]. 
This immediately allows us to extract the spin-dependent, excited-state decay rates: $\tau_0 = 1/(\Gamma_E + r\Gamma_{\mathrm{ISC}}) = 1.18(9)$~ns (from $|m_s = 0 \rangle$) and $\tau_1 = 1/(\Gamma_E + \Gamma_{\mathrm{ISC}})=0.47(5)$~ns (from  $|m_s = \pm1 \rangle$)~\cite{baber,jacques}. 

\begin{figure}
    \centering
    \includegraphics[width=\columnwidth]{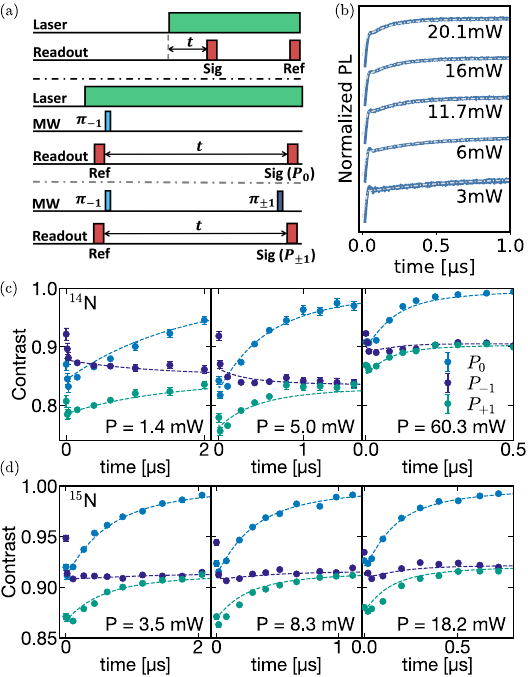}
    \caption{(a) Pulse sequences for power-dependent photoluminescence time traces (top) and spin-resolved polarization dynamics measurements (bottom). (b) Photoluminscence time traces from an unpolarized initial state at different optical powers. The dashed lines represent the results obtained from our semiclassical model with the rates in Table~\ref{tab:table1}.
    (c,d) Spin-resolved polarization dynamics measurements at different optical powers for (c) \nfour~ and (d) \nfive~ defect isotopes. 
    The dashed lines represent the results obtained from our semiclassical model.}
    \label{fig3}
\end{figure}

Next, we turn to a series of measurements aimed at carefully probing the power-dependence of the \vbm~polarization dynamics. 
Consistent with both \emph{ab initio} calculations and prior experiments~\cite{ivady2020, reimers2020, baber, whitefield, jacques}, our measurements above indicate an extremely fast intersystem crossing rate, $\Gamma_{\mathrm{ISC}}$.
Thus, the saturation time-scale of the polarization dynamics depends primarily on  the excitation rate, $\Gamma_P$, and the singlet lifetime, $\Gamma_s$.
At the lowest powers [Fig.~\ref{fig3}(b)], after a rapid initial rise, the photoluminescence exhibits a slow, power-dependent approach to its final saturation value. 
Meanwhile, at the highest powers, the time-scale at which the photoluminescence saturates starts to become independent of $\Gamma_P$, instead becoming constrained by the singlet lifetime~\cite{supp}. 

To further refine our estimates on the branching ratios, $\{r,k\}$, we perform an extensive set of measurements on the power-dependent, spin-resolved polarization dynamics for defects in both conventional (\nfour) hBN  and isotopically-purified (\nfive) hBN~\cite{supp}.
In particular, we characterize how the \vbm~approaches its polarized steady-state by measuring the effective population, $\{P_0, P_{\pm 1}\}$, in each spin sublevel as a function of time~\cite{supp}.
We initialize the ensemble in the $|m_s = -1\rangle$ ground state by applying a microwave $\pi$-pulse [Fig.~\ref{fig3}(a)]. 
Throughout the experiment, we apply green laser excitation to continuously polarize the system. 
The photoluminescence effectively tracks the population of the $|m_s = 0 \rangle$ state as a function of time; thus, one can measure the population of any spin sublevel, by simply moving it, with an additional $\pi$-pulse, to $|m_s = 0 \rangle$ before readout [Fig.~\ref{fig3}(a)]. 

There are two key features in the data [Fig.~\ref{fig3}(c,d)]. 
First, given that the initial state is in $|m_s = -1 \rangle$ and that polarization is toward the $|m_s = 0 \rangle$ state, one might expect minimal dynamics for the $|m_s = +1 \rangle$ state.
However, the branching ratio, $k$, drives population into the $|m_s = +1 \rangle$ state via the singlet.
Thus, the observed population transfer into the $|m_s = +1 \rangle$ state provides direct information about $k$.
Second, we find that the behavior  of the $|m_s = -1 \rangle$ state is non-monotonic at high laser powers.
This arises from a competition between the population dynamics driven by the ground-state microwave pulse and the laser excitation~\cite{supp}. 

Combining all of the aforementioned measurements, let us now extract the full set of parameters within the semiclassical model. 
We simultaneously fit the photoluminescence time traces [Fig.~\ref{fig3}(b)] and the spin-resolved polarization dynamics [Fig.~\ref{fig3}(c,d)], across all laser powers and both nuclear spin isotopes, under the constraints imposed by $\tau_0$ and $\tau_1$.
This yields the transition rates shown in Table~\ref{tab:table1}.

A few remarks are in order.
First, we extract an extremely small ISC branching ratio, bounded by $r \lesssim 0.04$.
This contrasts with most previous works~\cite{baber, jacques}, which have taken $r\sim0.5$ by directly utilizing timescales from excited-state spectroscopy and assuming a slow spin-conserving decay, $\Gamma_E$~\cite{ivady2020,reimers2020}.
Our estimate of $r$ yields two immediate implications: (i) it suggests that the \vbm~exhibits one of the best off-resonant polarization fidelities, with $p\gtrsim95\%$, for any defect and (ii) it predicts a very high ODMR contrast for experimentally accessible laser intensities~[Fig.~\ref{fig1}(d,e)]. 
This latter point helps to explain previous observations of ODMR contrasts up to $\sim46\%$~\cite{tongcang}, and suggests the ability to achieve readout fidelities near unity.
Second, our extracted spin-conserving decay rate, $\Gamma_E = 849~$MHz, is four orders of magnitude higher than  \textit{ab initio} predictions (Table~\ref{tab:table1}).
It is also in tension with the dimness of \vbm~ compared to other color centers. 
Similar observations have led to the speculation that there exists an additional, non-radiative, spin-conserving decay channel from ${}^3 E^{''}$ to ${}^3 A_2^{'}$~\cite{whitefield,jacques}.
We note that uncertainties about defect densities and the populations of various charge states could also contribute to this discrepancy, which remains an important open question.

\emph{Nuclear spin polarization dynamics}---Having carefully built  and characterized a model for the electronic spin dynamics, we are now in a position to incorporate additional ingredients, ranging from the \vbm~defects' intrinsic dipolar interactions to the effects of charge ionization and nuclear spins.
The latter is particularly important. 
Since the isotopes of both boron and nitrogen exhibit non-zero nuclear spin, the \vbm~is completely surrounded by a spin bath. 
On the one hand, this nuclear spin bath leads to relatively short decoherence time-scales~\cite{gottscholl2021odmr, murzakhanov2022, lee2022, haykal2022, gong2023}.
On the other hand, the ability to control nearby nuclear spins could also render them a potential resource, e.g. as a quantum register~\cite{dutt2007, ruskuc2022} or for quantum simulation~\cite{wang2015}.

\begin{figure}
    \centering
    \includegraphics[width=1.0\columnwidth]{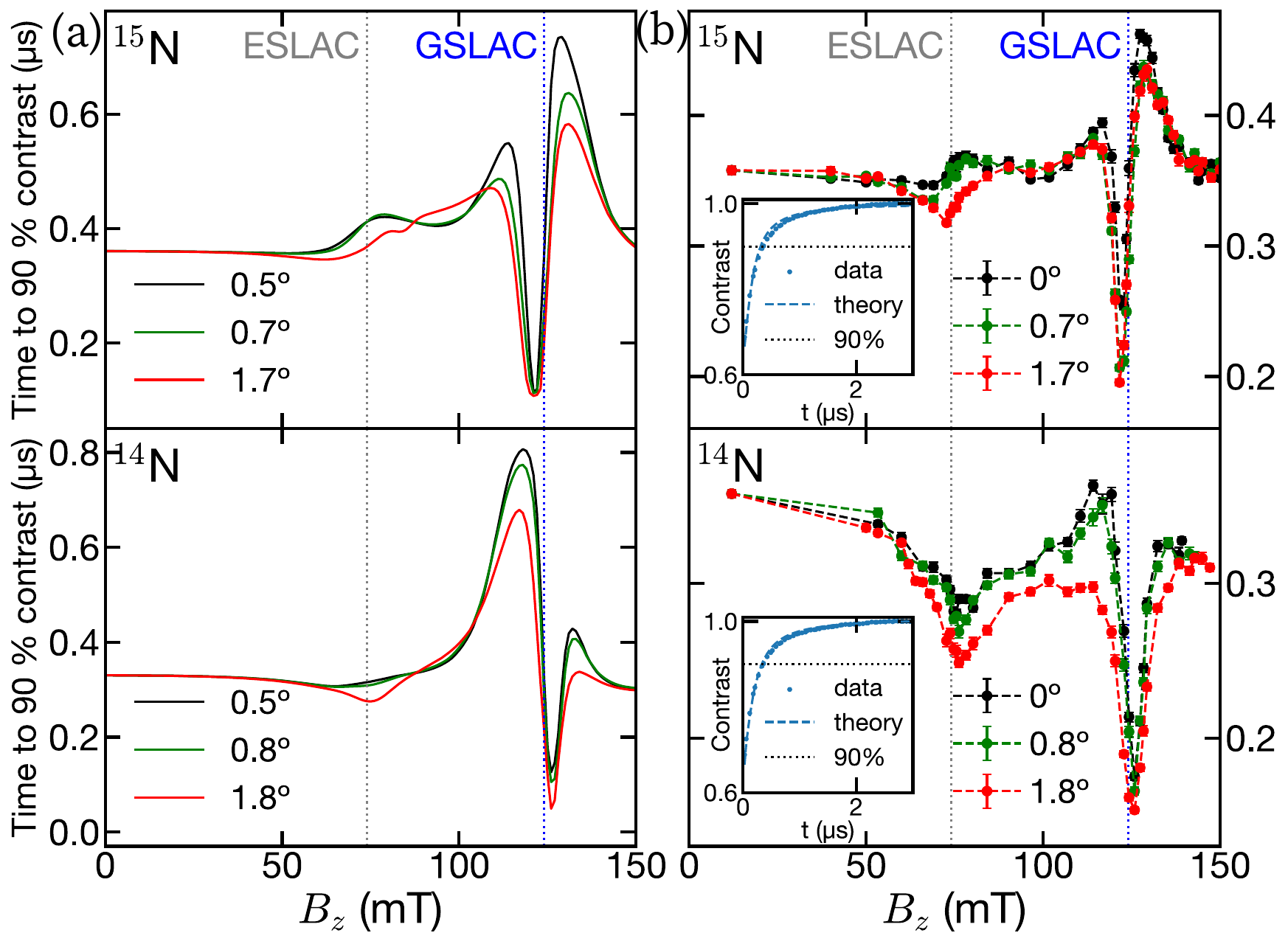}
    \caption{(a) Linbladian simulations for the electronic  spin polarization timescale as a function of both the magnetic field strength and orientation for \nfive~(top) and \nfour~(bottom) defect isotopes. 
    The numerics combine our semiclassical model with the rates from Table~\ref{tab:table1} using $r = 0$ and the coherent Hamiltonian dynamics from Eqn.~\ref{eqn_ham}, where $D_{\mathrm{gs}} = 2 \pi \times 3.48\,\text{GHz}$ and $D_{\mathrm{es}} = 2 \pi \times 2.1\,\text{GHz}$ are the zero-field splittings in the ground and excited states, respectively. 
    The electronic spin gyromagnetic ratio is $\gamma_e = 2 \pi \times 28\,\text{MHz/mT}$, while the nuclear spin gyromagnetic ratios are $\gamma_n^{{}^{14}\text{N}} = 2 \pi \times 3.076\,\text{kHz/mT}$ for \nfour~ and $\gamma_n^{{}^{15}\text{N}} = - 2 \pi \times 4.315\,\text{kHz/mT}$ for \nfive~ nitrogen nuclei. 
    (b) Corresponding experimental measurements of the electronic spin polarization timescale as a function of magnetic field strength and orientation, with an angular resolution of $1^{\circ}$~\cite{supp}.
    Inset: Photoluminescence time trace measured at $B_z = 12~$mT, compared to the numerical simulation.}
    \label{fig4}
\end{figure}

To this end, we now turn to characterizing the coupled dynamics of the \vbm~electronic and nuclear degrees of freedom.
The Hamiltonian governing the system is given by: 
\begin{equation}\label{eqn_ham}
H = D S^2_z + \gamma_e \mathbf{B} \cdot \mathbf{S} - \sum_i \gamma_n^i \mathbf{B} \cdot \mathbf{I}^i + \sum_i \mathbf{S} \mathbf{A}^i \mathbf{I}^i,
\end{equation}
where $\mathbf{S}$ is the electronic spin-1 operator (with a quantization axis out of plane) and $\mathbf{I}$ is the nuclear spin operator with $I=1$ for \nfour~and $I=1/2$ for \nfive. 
Here, $\mathbf{B}$ denotes the external magnetic field,  $\gamma_e$,  $\gamma_n$ are the electronic and nuclear spin gyromagnetic ratios,  $i$ indexes the surrounding nuclear spins (where we will specifically consider the three adjacent nitrogen atoms), and  $\mathbf{A}$ corresponds to the hyperfine interaction between the electronic and nuclear spins. 

For generic magnetic fields, the dominant effect of the hyperfine interaction is provided by the diagonal $S_z A^i_{zz} I^i_z$ term, leading to multiple peaks in the ODMR spectrum corresponding to different nuclear spin configurations~[Fig.~\ref{fig1}(b)].
However, much recent attention has focused on the ground- (GSLAC) and excited-state level anticrossings (ESLAC), where a magnetic field causes $|m_s=-1 \rangle$  to become degenerate with $|m_s = 0 \rangle$  (in either the ground or excited-state manifolds). 
At these fields, the hyperfine interaction can resonantly transfer polarization from the electronic spin to the nuclear spins, leading to asymmetry in the relative weights of the ODMR peaks~[insets, Fig.~\ref{fig1}(b)]~\cite{gao2022, gong2024}.
Understanding the dynamics of this polarization process as well as its dependence on the nuclear spin isotope remains an open question.

To this end, we utilize a  Linbladian simulation that combines our semiclassical model with the coherent Hamiltonian dynamics from Eqn.~(\ref{eqn_ham}), to predict the full polarization dynamics of the coupled electronic-nuclear spin system. 
We investigate these dynamics for both nuclear spin isotopes as a function of the external magnetic field strength and tilt [Fig.~\ref{fig4}(a)]; the strength controls the system's proximity to GSLAC and ESLAC, while the tilt tunes a direct coupling between the
$|m_s = 0 \rangle$ and $|m_s= \pm1 \rangle$ states. 
As a simple proxy for characterizing these dynamics, we plot the time-scale for the photoluminescence to cross a particular threshold [Fig.~\ref{fig4}(a)].

Our simulations predict three qualitative effects.
One expects the polarization time-scale to exhibit a peak near both GSLAC and ESLAC.
Intuitively, this arises from the  transfer of polarization to the nuclear spins, which slows down the overall polarization dynamics. 
Such a peak is indeed partially observed at GSLAC [Fig.~\ref{fig4}(a)], but is overwhelmed by a sharper dip at both level anticrossings. 
This dip results from a coherent mixing between the different electronic spin sublevels, which yields a direct channel for polarization into the $|m_s = 0\rangle$ state without needing to traverse the intersystem crossing. 
Two effects contribute to this mixing: (i) an off-axis magnetic field directly couples the sublevels and (ii) the hyperfine interaction yields an additional indirect, nuclear-spin mediated coupling. 
Finally, our numerics suggest that the positions of these two features (i.e.~the peak and dip in time-scale) are reversed for the two isotopes: For \nfour, the peak is centered slightly to the left of GSLAC, while the dip is centered slightly to the right; for \nfive, the opposite is true [Fig.~\ref{fig4}(a)].
Interestingly, this owes to a sign difference in the  gyromagnetic ratios of the two isotopes, which in turn, inverts the sign of all of the hyperfine couplings.

We now turn to the analogous experiments.
As shown in [Fig.~\ref{fig4}(b)], starting from an unpolarized initial state, we measure the polarization dynamics as a function of magnetic-field strength and tilt, for both isotopes.
All three of the predicted features are indeed borne out in our data.

Our work opens the door to a number of intriguing future directions. 
First, the predicted near-unity electronic spin polarization, suggests the use of \vbm~ensembles as a possible room-temperature quantum simulation platform.
Indeed, by starting from the polarized state and adiabatically ramping down a pinning field in the same direction, one could explore the formation of dipolar spin glasses~\cite{rosenbaum1991dipolar,reich1990dipolar,alonso2015low}.
Second, the high degree of electronic spin polarization also suggests the ability to generate a correspondingly  large nuclear spin polarization.
If achievable, this would significantly reduce inhomogeneous broadening and enable the nuclear spins to become viable quantum registers themselves~\cite{dutt2007quantum,ruskuc2022nuclear,van2024mapping}. 
Finally, our methods and approach can naturally be applied to the emerging class of carbon-based defects in hBN, in order to characterize both their polarization and nuclear spin dynamics~\cite{stern2022, stern2024}.

\emph{Acknowledgements}---We gratefully acknowledge the insight of and discussions with H. Park, I. Luxmoore, A. McClelland, X. Liao, N. Sinclair, A. Zibrov, P. Zhou, and Y. Zhu.
This work was supported in part by multiple grants from the  U.S. Department of Energy,  Office of Science, Office of Basic Energy Sciences, including via the Division of Chemical Sciences, Geosciences, and Biosciences at LBNL under Contract No.~DE-AC02-05CH11231 and via BES grant No.~DE-SC0019241. The support for hexagonal boron nitride crystal growth was provided by the Office of Naval Research, Award No.~N00014-22-1-2582.
P.~Kim and Z.~Hao acknowledge support from the ONR MURI (N00014-21-1-2537).
W.~Lee acknowledges support from the  Center for Ultracold Atoms (an NSF Physics Frontier
Center).
N. Y. Yao acknowledges support from a  Brown Investigator award.

\nocite{*}
\bibliography{ms}


\end{document}


\title{Supplemental Material for Intrinsic high-fidelity spin polarization of charged vacancies in hexagonal boron nitride}
\date{\today}
\maketitle
\section{hBN device fabrication and experimental setup}

\subsection{hBN device fabrication}

We compare the nuclear spin dynamics of different nitrogen isotopes \nfour~and \nfive. The natural abundant $\mathrm{hBN}_\mathrm{nat}$ consists of $99.6\%/0.4\%$ for \nfour/\nfive~ and $20\%/80\%$ for ${}^{10}\mathrm{B}/{}^{11}\mathrm{B}$. The $\mathrm{h}^{10}\mathrm{B}^{15}\mathrm{N}$ crystals were grown from a molten mixture of 48 wt.\% nickel (99.99\%), 48 wt\% chromium (99.99\%), and 4 wt.\% of isotopically enriched elemental boron-10 (98\% $^{10}\mathrm{B}$, 99.999\% pure) heated to a maximum temperature of 1550 °C in a horizontal alumina tube furnace. The crystal growth was performed under 860 torr of static ultra high-purity hydrogen (99.999\%) gas and isotopically enriched nitrogen gas (97\% $^{15}\mathrm{N}_2$, 99.999\% pure). This solution was then slowly heated to 1500 °C at 1 °C/h to precipitate crystal formation. After the ingot was cooled to room temperature, it was removed from the furnace and freed from the crucible. The hBN crystals were removed by pressing thermal release tape against the surface of the ingot and slowly lifting off. The tape was heated to deactivate the adhesive and the crystals were transferred using acetone and a paint brush. More details on growing hBN with specific boron and nitrogen isotopes were reported in Refs.~\cite{liu2018,janzen2024}.

We exfoliate the hBN crystal into thin flakes using tape, with the $\mathrm{hBN}_\mathrm{nat}$ flake having a thickness of $60 \pm 10~$nm and the $\mathrm{h}^{10}\mathrm{B}^{15}\mathrm{N}$ flake having a thickness of $84 \pm 10~$nm. The \vbm~ defects are created via $\mathrm{He}^{+}$ ion implantation with an energy $3~$keV and a dosage of $1~\mathrm{ion}/\mathrm{nm}^2$ at CuttingEdge Ions LLC. These results in an estimated \vbm 
~concentration of around $150~$ppm \cite{gong2023}. For microwave delivery, we use a coplanar waveguide with a $50~\mu$m central line fabricated onto a $400~\mu$m thick sapphire wafer [Fig.~1(c) in the main text]. After ion implantation, we transfer the hBN flakes onto the waveguide using polycarbonate (PC) stamps. Once the transfer is complete, we clean the residual PC films by immersing the waveguide in chloroform.

\subsection{Experimental setup}
In our experiment, we use a home-built confocal microscope setup. A 532 nm laser (Lighthouse Photonics Sprout-H) is used for initialization and readout of \vbm. The laser beam is focused to a diffraction-limited spot with a diameter of $\sim 0.7~\mu$m using an objective (Olympus LCPlanFL 40x Ph2). The objective lens is mounted on a piezo scanner (Edmund optics 85-008) for vertical positioning control of the laser beam relative to the hBN flake. For the lateral control, the laser beam is scanned with a X-Y galvanometer (Thorlabs GVS212). The laser beam is shuttered with an acousto-optic modulator (AOM, G\&H AOMO 3110-120) in a double-pass configuration to achieve $>10^4/1$ on/off ratio. To detect the photoluminescence of \vbm, we use a dichroic mirror (Semrock LM01-552-25) and a 532 nm long-pass filter (Semrock BLP01-594R-25) to separate the collection path from the laser beam. The photoluminescence is detected by a single photon counting module (Excelitas SPCM-AQRH-63-FC).  

To apply magnetic fields, we use a set of three electromagnet coils to provide $x,y$, and $z$ field control, with the $z$-axis is defined as the $c$-axis of the hBN lattice. We calibrate the vertical and transverse fields by systematically measuring the resonance frequencies of optically detectable magnetic resonance (ODMR) as a function of the applied magnetic field magnitude. An external $B_z$ field separates the $|m_s = \pm 1 \rangle$ states with a splitting of $f_{+1} - f_{-1} = 2\gamma_e B_z$, where $\gamma_e = (2 \pi) \times 28~$MHz/mT is the gyromagnetic ratio of the electronic spin. In addition, the transverse magnetic field $B_t = \sqrt{B_x^2 + B_y^2}$ shifts the center frequency of the $|m_s = \pm 1 \rangle$ resonance peaks, $f_c = (f_{+1} + f_{-1})/2$ by $\Delta f_c = 0.75 \gamma_e^2 B_t ^2 \{ 1/(D_{gs} - \gamma_e B_z) + 1/(D_{gs} + \gamma_e B_z)\} $ where $D_{gs} = 2 \pi \times 3.48\,\text{GHz}$ is the electronic zero-field splitting in the ground state. For fine calibration of the transverse field, we compare the \vbm~ polarization timescales [Fig.~4 in the main text] and photoluminescence as a function of $B_t$ magnitude around ground-(GSLAC) and excited-state level anti-crossings (ESLAC) [Fig.~\ref{fig:supp_nuc}]. This allows for the calibration of the $B_z$ field with a tilt angle $\theta = \mathrm{arctan}(B_t / B_z)$, achieving an angular resolution of approximately $1^\circ$. 

In the presence of an external $B_z$ field, we can drive transitions between the $|m_s = 0 \rangle$ and the $|m_s = \pm 1 \rangle$ states using microwave field pulses. For this purpose, we use a microwave generator (Stanford Research SG386) and an arbitrary wave generator (AWG, Chase Scientific Wavepond DAx22000). Specifically, we utilize the in-phase/quadrature (IQ) modulator function of the microwave generator to mix its signal with that of the AWG and generate a resonant microwave pulse. The microwave generator can produce high-frequency signals ranging from 1 to $6~$GHz. The AWG, with a sampling rate of $2~$GHz, provides a temporal resolution of $0.5~$ns, enabling the generation of a sufficiently short microwave pulses for high fidelity. To eliminate any noise leakage, the microwave pulse is shuttered with a switch (Minicircuits ZASWA-2-50DRA+). We use a microwave amplifier (Minicircuits ZHL-50W-63+) to amplify the microwave signal and deliver it to the coplanar waveguide. The pulse sequence of the AOM control, microwave pulses, and SPCM gate is produced by a multi-channel pulse generator (SpinCore PulseBlasterESR-PRO 500).

\section{Spin-dependent excited spectroscopy: experimental details}

\begin{figure}
    \centering
    \includegraphics[width=0.8\textwidth]{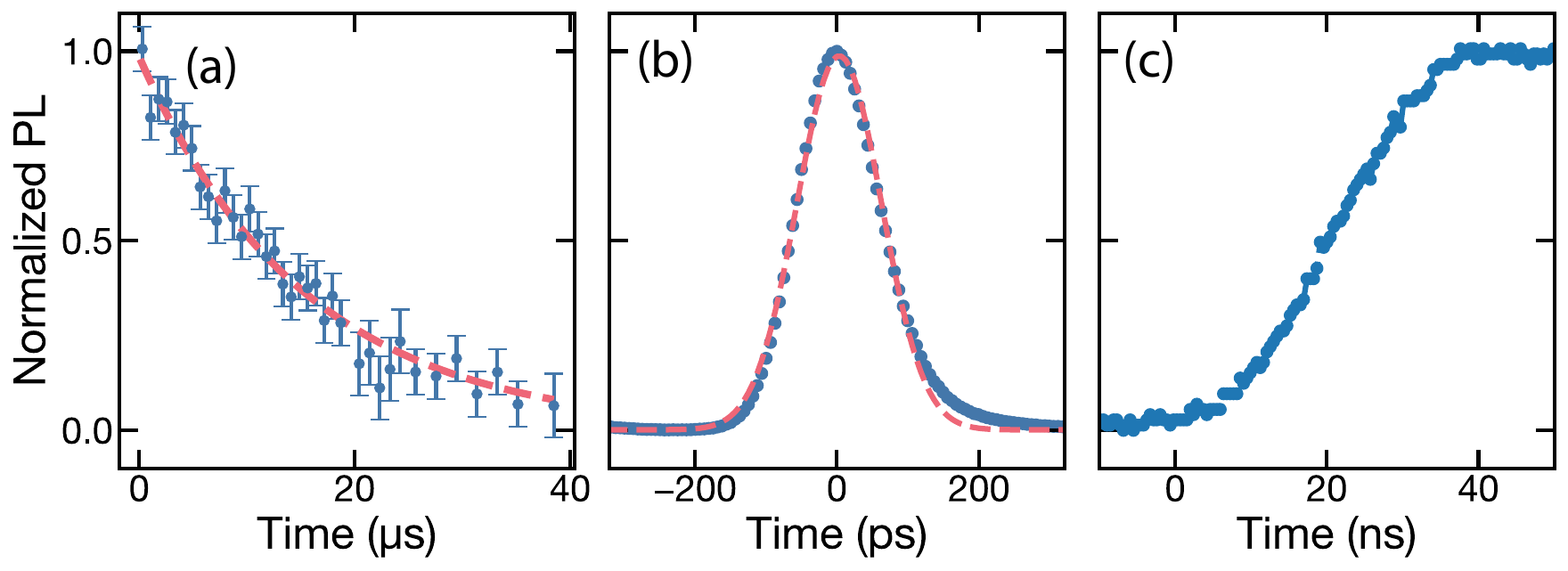}
    \caption{(a) Spin relaxation curve. The dashed line represents a fit to a single exponential decay with $T_1 = 15.4(7)~$µs. (b) Instrument response function fitted with a Gaussian distribution of $\mathrm{FWHM}_\mathrm{IRF} = 140(1)~$ps. (c) Temporal response of the laser pulse by detecting the light reflected by a Spectralon diffuse reflectance.}
    \label{fig:IRF_AOM_T1}
\end{figure}

To measure the spin-dependent excited-state lifetime, we perform time-resolved photoluminescence measurements [Fig.~2(b) in the main text]. We use a broadband supercontinuum laser (SuperK VARIA) with adjustable wavelength, bandwidth, repetition rate, and a pulse width of approximately 5 ps. For \vbm~excitation, we set the wavelength to 532 nm with a bandwidth of 50 nm. To detect the photoluminescence signal of \vbm~ensembles, we use a fast single-photon detector (IDQ ID100) with a timing resolution of 40 ps. Time-resolved photoluminescence measurements are conducted using a time-tagging module (Swabian Time Tagger Ultra).

The pulse sequence of the excited-state spectroscopy is as follows. A 10 µs-long laser pulse train is applied to optically polarize the \vbm~to the ground $m_s$ = 0 state. The laser pulse has a repetition frequency of 39 MHz, with each pulse having a width of 5 ps. For the photoluminescence readout, another laser pulse train is applied after 1 µs, which is much shorter than the measured $T_1$ = 15 µs [Fig.~\ref{fig:IRF_AOM_T1}(a)]. The photoluminescence decay signal is averaged over a 1 µs duration, during which the initial 39 repetitions of the laser pulses occur. For the differential measurements, we apply a $\pi$-pulse of 26 ns length to initialize the \vbm~to the ground $m_s$ = -1 state and repeat the same readout process. 

Several factors contribute to the instrumental response function, which limits the timing resolution of the time-resolved photoluminescence measurements. These factors include the finite laser pulse width, the time jitter of the single-photon detector and time-tagging module, and other electronic noises. Fig.~\ref{fig:IRF_AOM_T1}(b) shows the measurement of our instrumental response function, which is fit to a Gaussian function with a broadening of $\mathrm{FWHM}_\mathrm{IRF} = 140(1)~$ps. Consequently, we fit the measured photoluminescence signal to a bi-exponential form convolved with a Gaussian of 140 ps width [Fig. 2(b) in the main text].


\section{Simulations of the semiclassical rate model}

As described in Fig. 2(a) in the main text, our semiclassical rate model describes the populations of seven quantum levels: the spin-1 ground state manifold $\left| g_{0, \pm 1} \right>$, the spin-1 excited state manifold $\left| e_{0, \pm 1} \right>$, and the metastable spin-0 singlet level, $\left| s \right>$. In the basis $\left\{ g_{+1}, g_{0}, g_{-1}, e_{+1}, e_{0}, e_{-1}, s \right\}$, the dynamics are governed by the relation
\begin{align}
\resizebox{\textwidth}{!}{$
\dot{\mathbf{P}} = \begin{pmatrix}
-\Gamma_P - \gamma_1/2 & \gamma_1/2 & 0 & \Gamma_E & 0 & 0 & k \Gamma_s \\
\gamma_1/2 & -\Gamma_P - \gamma_1 & \gamma_1/2 & 0 & \Gamma_E & 0 & \Gamma_s \\
0 & \gamma_1/2 & -\Gamma_P - \gamma_1/2 & 0 & 0 & \Gamma_E & k \Gamma_s \\
\Gamma_P & 0 & 0 & -\Gamma_E - \Gamma_\mathrm{ISC} - \gamma_1/2 & \gamma_1/2 & 0 & 0 \\
0 & \Gamma_P & 0 & \gamma_1/2 & -\Gamma_E - r \Gamma_\mathrm{ISC} - \gamma_1 & \gamma_1/2 & 0 \\
0 & 0 & \Gamma_P & 0 & \gamma_1/2 & -\Gamma_E - \Gamma_\mathrm{ISC} - \gamma_1/2 & 0 \\
0 & 0 & 0 & \Gamma_\mathrm{ISC} & r \Gamma_\mathrm{ISC} & \Gamma_\mathrm{ISC} & -\left( 1 + 2k \right) \Gamma_s
\end{pmatrix}
\mathbf{P},
$}
\end{align}
where $\mathbf{P}$ represents the populations in the seven quantum levels. Here, $\Gamma_P$, $\Gamma_E$, $\Gamma_\mathrm{ISC}$, $\Gamma_s$, $r$, and $k$ are the transition rates defined in the main text, while $\gamma_1$ denotes the spin depolarization rate described by $T_1=1/\gamma_1=15\,\text{µs}$ [Fig.~\ref{fig:IRF_AOM_T1}(a)], which we assume is the same for both the ground and excited states. For estimation of the degree of spin polarization [Fig.~1(d) in the main text], we calculate the polarization $p = P_{0} / (P_0 + P_{+1} + P_{-1})$ as a function of time, where $P_{0,\pm 1}$ represents the populations in each spin sublevel of $|m_s = 0, \pm 1\rangle$, combining both ground and excited states. For calculations of ODMR contrast from microwave mixing [Fig.~1(e) in the main text], we include an additional term that transfers populations between $g_0$ and $g_{\pm 1}$ at a rate of $\Omega_{\pm 1}$: we find that this incoherent model is nearly identical to the effect of a coherent driving term (in a full Lindbladian treatment) in the steady state, and thus suffices for our calculations.

We note that the laser incident on our sample has an approximately Gaussian beam intensity profile. This modifies our transition rate model such that $\Gamma_P$ is no longer a uniform distribution but instead a two-dimensional Gaussian; in our numerics, we discretize this Gaussian distribution into five approximate sectors, with modified intensities as a factor of $I = \left( 0.923, 0.485, 0.134, 0.0194, 0.00148 \right) \cdot I_0$, where $I_0$ is the intensity at the center of the beam. We find that this level of precision fully captures the effect of a Gaussian beam.

\subsection{Experimental data fitting}

With this semiclassical rate model, we can now simulate the dynamics of \vbm~and extract transition rates by fitting to experimental data. As described in the main text, we use constraints on $\tau_0$ and $\tau_1$ provided by excited state spectroscopy [Fig.~2(b) in the main text] and simultaneously fit our measurements of the photoluminescence time traces and spin-resolved polarization dynamics.

A few details are worth mentioning. Most importantly, as emphasized in the main text, we consider the effect of other emitters in hBN by allowing our fits to include a power-dependent constant background fluorescence in each dataset we analyze. While this implies that figures of absolute photoluminescence cannot be directly compared across different datasets, the most important features---timescales, and comparisons of photoluminescence at equal laser power---are unaffected or become more robust.

In addition, we account for various experimental details in all of our measurements. In our photoluminescence time trace experiment, we measure the response function of the acousto-optical modulator (AOM) that controls the initialization of our laser to be approximately linear with a timescale of $40\,\text{ns}$ [Fig.~\ref{fig:IRF_AOM_T1}(c)], which we then integrate into our simulations. In our spin-resolved polarization dynamics experiments, there are multiple features to take into account. First, our measurements are averaged for $125\,\text{ns}$ at the end of each sequence---this is necessary in order to properly reflect the change in ground-state spin populations after the end of the last $\pi$-pulse. Our simulations reflect this averaging time accordingly. Second, as expressed in the main text, there is a competition between the laser-driven and microwave-driven dynamics in the duration of each $\pi$-pulse, which leads to a proportion of the original population to fail to be driven to the desired state. We model this phenomenon in terms of an effective fidelity $f$, such that a fraction $f$ of the population is successfully transferred in the $\pi$-pulse. Third, since our pulse sequences [Fig.~3(a) in the main text] are not differential measurements, we cannot independently measure the populations in the $|m_s = 0, \pm 1\rangle$ ground states. However, we can probe the effective populations $\{P_0, P_{\pm 1}\}$ by noting that $m_s = 0$ state gives a much higher photoluminescence signal compared to the $m_s = \pm 1$ states due to the small branching ratio $r$ [Fig. 2(a) in the main text]. 

\begin{figure}
    \centering
    \includegraphics[width=\textwidth]{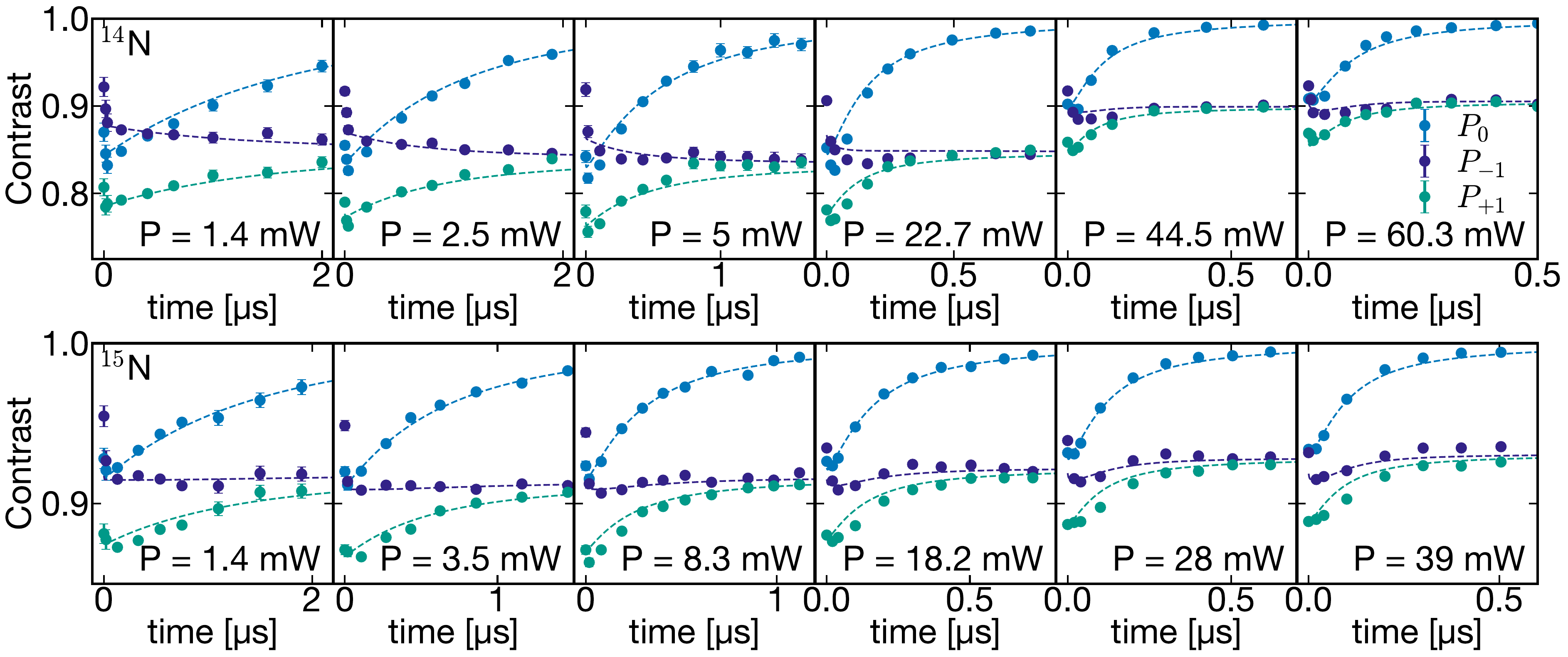}
    \caption{Full spin-resolved polarization dynamics datasets used for fitting, for \nfour~(top) and \nfive~(bottom) defect isotopes, compared to simulations with best fit parameters [Table I in the main text].}
    \label{fig:supp_bd}
\end{figure}

The results of our spin-resolved polarization dynamics measurements, as well as comparisons with the simultaneous best fit, are summarized in Fig.~3(c,d) of the main text, with the full data plotted in Fig.~\ref{fig:supp_bd}. As explained in the main text, the competition between laser and microwave dynamics is the cause of the non-monotonicity in the $P_{-1}$ signal (due to effectively mixing with the $P_{0}$ signal), as well as the fact that the $P_{+1}$ signal starts at a lower value than equilibrium. Furthermore, this non-monotonicity is more evident at higher laser powers, making the higher power datasets better fit by lower effective fidelities $f$. To account for this effect, we allow $f$ to decrease linearly with laser power, or $\Gamma_P$, and allow this dependence to be determined as a fit parameter. We note that this simplistic model does not fully capture the size of the monotonicity for all powers, or other effects such as an observed difference in the effective fidelities of high-power \nfive~datasets; however, it suffices to explain all qualitative features of the data, while more complex models of $f$ would necessarily rely on additional assumptions or fine-tuning and run the risk of overfitting.

The best fit results of the universal rates $\left\{ \Gamma_\mathrm{ISC}, \Gamma_E, \Gamma_s, r, k \right\}$ are given in Table I in the main text. The remaining non-universal fit parameters, which are dependent on our specific hBN sample and optical setup, are as follows: $\alpha = \left\{ 1.319, 1.649, 1.864 \right\} \,\text{MHz/mW}$ for our \nfour~and \nfive~samples in the spin-resolved polarization dynamics measurements, and our sample in the photoluminescence time trace measurement, respectively; $f = \left\{ 0.615-0.00341 \left( \Gamma_P/\text{MHz} \right), 0.499-0.00307 \left( \Gamma_P/\text{MHz} \right) \right\}$ for our \nfour~and \nfive~samples respectively.

We note that, although our best fit has $r=0$ exactly, values of $r$ up to $\sim 0.04$ are reasonable; achievable levels of polarization and ODMR contrast vary as a result. Therefore, we obtain best fits as a function of varying $r$: with a relatively small $\Gamma_P = 20\,\text{MHz}$, $r=0.04$ corresponds to polarization $p \sim 95\%$ and ODMR contrast of up to $40\%$ with high microwave mixing. In Fig.~1(d,e) in the main text, we use the best-fit results of the rates given in Table I in the main text and $r = 0$. It is also important to mention that Appendix C of ref.~\cite{jacques} reports a relatively small $r = 0.027$ for the case where a direct non-radiative decay rate is included in their model.

\section{Lindbladian simulations}

To model the coherent effects of the \vbm~system, especially that of the surrounding nuclear spins, we utilize the Lindblad master equation formalism,
\begin{align}
\dot{\rho} = -\frac{i}{\hbar} \left[ H, \rho \right] + \sum_i \left( c_i \rho c_i^\dagger - \frac{1}{2} \left\{ c_i^\dagger c_i, \rho \right\} \right),
\end{align}
which describes the dynamics of the full density matrix $\rho$ under both Hamiltonian ($H$) and incoherent ($c_i$) dynamics for jump operators $c_i$. As detailed in Eq. 1 in the main text, the Hamiltonian of the system is given by
\begin{align}
H = D S^2_z + \gamma_e \mathbf{B} \cdot \mathbf{S} - \sum_i \gamma_n^i \mathbf{B} \cdot \mathbf{I}^i + \sum_i \mathbf{S} \mathbf{A}^i \mathbf{I}^i,
\end{align}
where $\mathbf{S}$ and $\mathbf{I}^i$ index the electronic and nuclear spin degrees of freedom, respectively. The values of the electronic zero-field splitting $D$ are $D_{gs} = 2 \pi \times 3.48\,\text{GHz}$ in the ground state and $D_{es} = 2 \pi \times 2.1\,\text{GHz}$ in the excited state, with the electronic gyromagnetic ratio of $\gamma_e = 2 \pi \times 28\,\text{MHz/mT}$. For the nuclear spins, we focus on the three adjacent \nfour~or \nfive~nitrogen nuclei, with $\gamma_n^{{}^{14}\text{N}} = 2 \pi \times 3.076\,\text{kHz/mT}$ and $\gamma_n^{{}^{15}\text{N}} = - 2 \pi \times 4.315\,\text{kHz/mT}$. For \nfour~with spin $I=1$, there is an additional quadrupole coupling term $Q_{zz} I_z^2$ for $Q_{zz}=2 \pi \times 0.79\,\text{MHz}$, though the effect of this term is almost negligible.

The hyperfine tensor $\mathbf{A}$, which couples the electronic and nuclear degrees of freedom, has been computed for \nfour~nuclear spins by \textit{ab initio} calculations for both the ground and excited states~\cite{ivady2020, gao2022}, which we utilize in our simulations. To obtain the couplings for \nfive~instead, one can simply multiply these values by the ratio $\gamma_n^{{}^{15}\text{N}}/\gamma_n^{{}^{14}\text{N}} \thickapprox -1.4$, thus inverting the sign of all couplings.

We now turn to describing the jump operators $c_i$ in the Lindbladian formalism. Most such operators correspond to the transition rates in the semiclassical model: the laser excitation $\Gamma_P$ is expressed as $c_P = \sqrt{\Gamma_P} \sum_{i \in 1,0,-1} \left| e_{i} \right> \left< g_{i} \right|$, while the spin-conserving decay $\Gamma_E$ is $c_E = \sqrt{\Gamma_E} \sum_{i \in 1,0,-1} \left| g_{i} \right> \left< e_{i} \right|$. The decay into the singlet can be written as the three jump operators $c_{\mathrm{ISC},0} = \sqrt{r \Gamma_\mathrm{ISC}} \left| s \right> \left< e_{0} \right|$ and $c_{\mathrm{ISC},\pm 1} = \sqrt{\Gamma_\mathrm{ISC}} \left| s \right> \left< e_{\pm 1} \right|$, while the decay out of the singlet is then $c_{s,0} = \sqrt{\Gamma_{s}} \left| g_0 \right> \left< s \right|$ and $c_{s,\pm 1} = \sqrt{k \Gamma_{s}} \left| g_{\pm 1} \right> \left< s \right|$. The $T_1$ decoherence operators for the electronic spin can be written in terms of the raising and lowering operators, with $c_{1,\pm} = \sqrt{\gamma_1/2} S_{\pm}$ for both the ground and excited states. In addition to these rates, we are now able to model the effect of dephasing ($T_2$) on the system. We can write $c_2 = \sqrt{\gamma_2} S_z$ for the ground and excited states, where $T_2 = 1/\gamma_2 = 62\,\text{ns}$~\cite{haykal2022}. In principle, we are also able to include $T_1$ and $T_2$ decoherence channels for the nuclear spins, but their values are unknown, and $T_1$ in particular is suspected to be high~\cite{gong2023,gong2024}.

\begin{figure}
    \centering
    \includegraphics[width=0.6\textwidth]{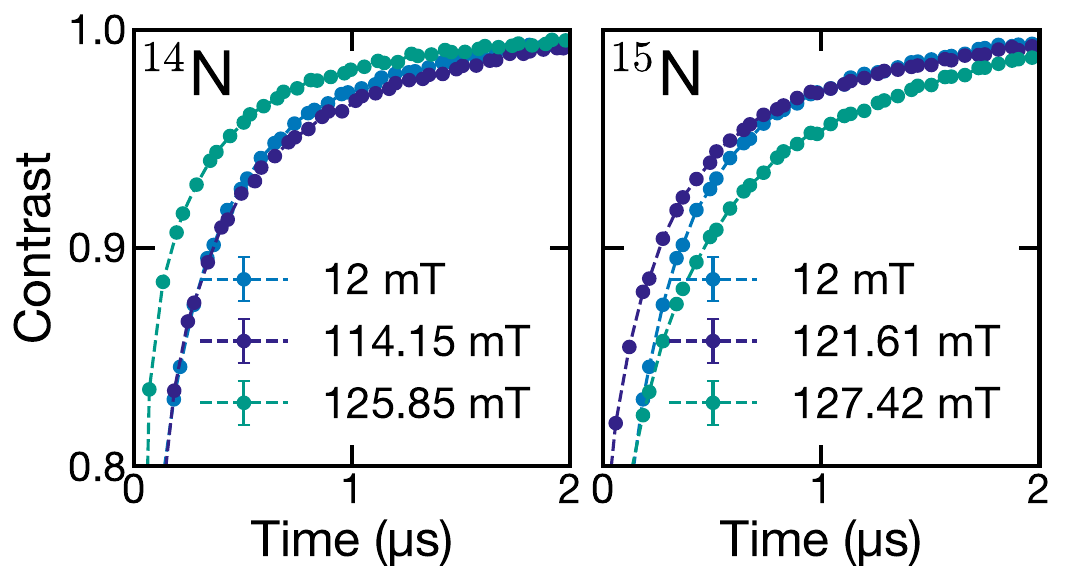}
    \caption{Photoluminescence time traces at different magnetic fields, including near GSLAC (124 mT) with a $B_z$ tilt of 0 degrees for \nfour~(left) and \nfive~(right) defect isotopes. }
    \label{fig:N15_N14_tscan}
\end{figure}

\begin{figure}
    \centering
    \includegraphics[width=0.7\textwidth]{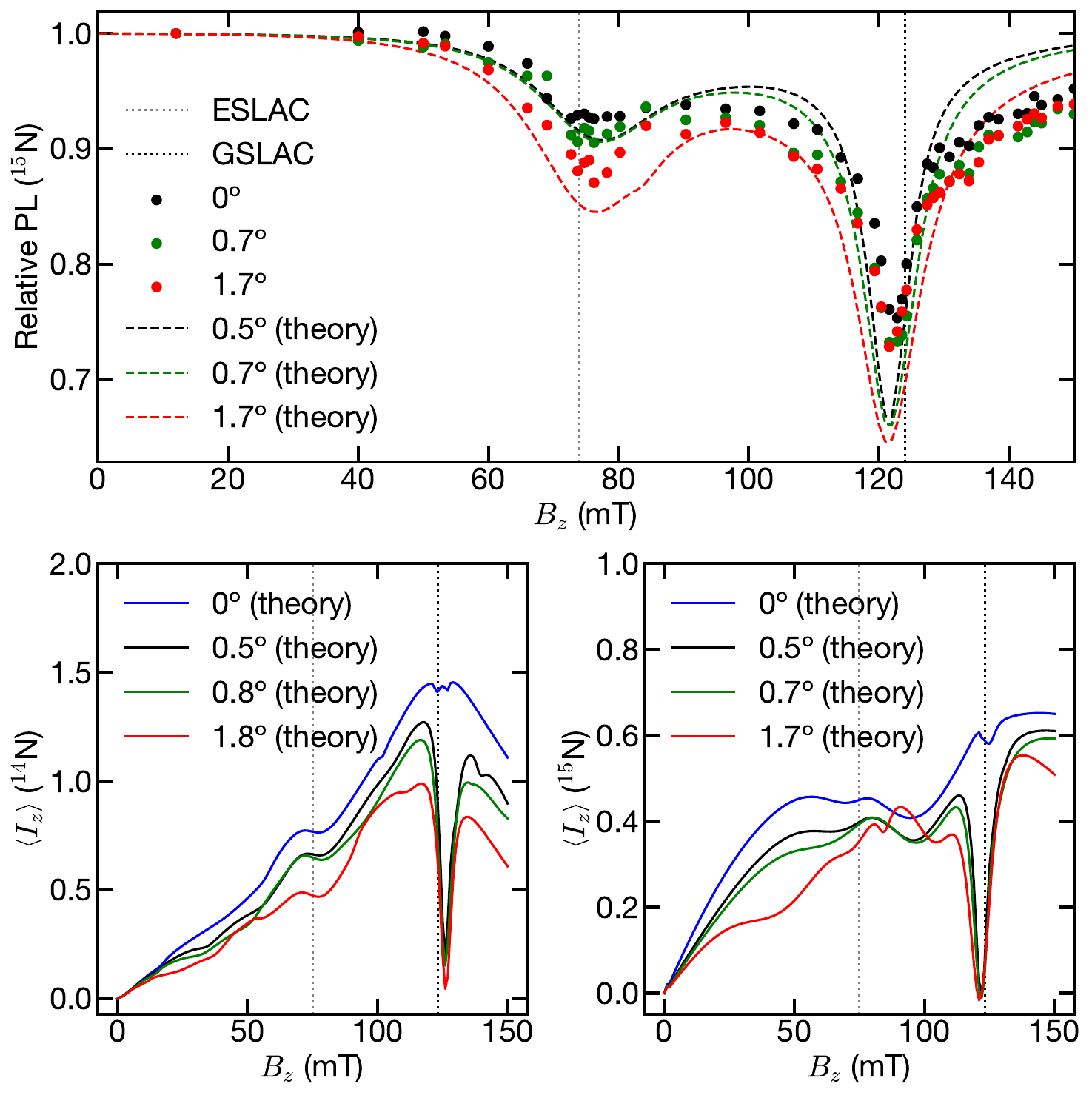}
    \caption{Top: Relative steady-state photoluminescence as a function of magnetic field strength $B_z$ and tilt angle $\theta$ for our \nfive~sample, with background photoluminescence calibrated using the dynamics measurements in Fig.~4 of the main text. Bottom: Theoretical calculations of nuclear spin polarization $\left< I_z \right>$ as a function of magnetic field for \nfour~(left) and \nfive~(right) defect isotopes.}
    \label{fig:supp_nuc}
\end{figure}

Equipped with this Lindbladian model, we are now able to perform simulations on the full dynamics of the system with the inclusion of nuclear spins, as discussed in the main text. To simulate the photoluminescence time trace during optical polarization [Fig.~4(a) in the main text], we calculate the total populations in the excited manifold as a function of time from the unpolarized initial state. To calculate the contrast, the reference photoluminescence is determined by the excited state populations at steady state. For comparison to the numerical simulation, we measure the photoluminescence time traces following the pulse sequence in [Fig.~3(a) in the main text] at different magnetic fields, ranging from 12 mT to 150 mT [Fig.~\ref{fig:N15_N14_tscan}]. 

To fit the simulated time trace to the experimental data, we include the background fluorescence $b$ from other emitters as a free parameter, using $\mathrm{contrast} = (\mathrm{sig} + b)/(\mathrm{ref} + b)$. This yields good agreement at low fields $B_z = 12~$mT [insets, Fig.~4(b) in the main text]. Assuming that the background fluorescence is independent of the fields, we then apply the same parameter $b$ to the simulated time traces at other $B_z$ fields and extract the polarization timescale [Fig.~4(a) in the main text]. 

In addition to dynamical quantities, we can also compute various steady state quantities, such as the photoluminescence as a function of magnetic field strength and magnetic field tilt, and the nuclear spin polarization $\left< I_z \right>$, as shown in Fig.~\ref{fig:supp_nuc}. For simplicity, we implement the magnetic field tilt angle $\theta$ as a small $B_x$ field in addition to the applied $B_z$, although the orientation of our magnetic field in the $x$-$y$ plane relative to the \vbm~$x$-axis (defined as the mirror symmetry plane of the hBN lattice) is unknown in our experiment.

\section{Spin-resolved polarization dynamics of NV centers}
The most distinctive feature of \vbm~is its small branching ratio $r\lesssim 0.04$ [Table I in the main text], which results in a high degree of spin polarization $p \gtrsim 95\%$ [Fig.~1(d) in the main text]. To validate our spin-resolved polarization dynamics measurement [Fig. 3(c,d) in the main text], we perform the same set of measurements with NV centers, which have well-characterized transition rates for the seven-level rate model [Table~\ref{tab:NV_rate}]. This approach allows us to verify whether we can determine the \( r \) value for the NV center in an unbiased manner.

\par
\begin{table}[h]
    \centering
    \begin{tabular}{|c|c|c|c|c|c|}
    \hline ~ & $\Gamma_{E}$ & $r\Gamma_\mathrm{ISC}$ & $\Gamma_\mathrm{ISC}$ & $\Gamma_{s}$ & $k\Gamma_{s}$ \\
    \hline Rate [MHz] & $66 \pm 5$ & $8 \pm 4$ & $53 \pm 7$ & $1.0 \pm 0.8$ & $0.7 \pm 0.5$ \\
    \hline
    \end{tabular}
    \caption{Transition rates for the seven-level rate model of NV centers~\cite{Robledo_2011, Tetienne_2012, Zu2021}.}
    \label{tab:NV_rate}
\end{table}
The NV sample used in our measurement consists of synthetic type-Ib 111-cut single crystal diamonds supplied by Sumitomo. The NV centers were created via a multi-step electron irradiation process. Initially, the samples were irradiated with electrons at an energy of 12 keV and a dosage of $4.5 \times 10^{11} \, \mathrm{cm}^{-2}$. This was followed by a second irradiation at 30 keV with a dosage of $1.35 \times 10^{11} \, \mathrm{cm}^{-2}$, generating vacancies. Post-irradiation, the diamonds were annealed in a vacuum environment at $10^{-6} \, \mathrm{Torr}$ and temperatures exceeding $800^\circ \mathrm{C}$. Simulations conducted using the SRIM software confirmed that the NV centers were predominantly distributed within a thin layer approximately 50 nm deep.\par

The critical parameter in our analysis is $r$, with most of its information derived from the $P_0$ signal, which probes the effective population of the $|m_s=0\rangle$ ground state.
Our fitting scheme closely follows the approach used for the \vbm~spin-resolved polarization dynamics measurements [Sec.~III]. Given the significant uncertainty in the branching ratio \( k \) and the singlet decay rate \( \Gamma_s \), which could affect the fitting parameter \( r \), we introduce an effective decay rate parameter \( \Gamma_s^* \equiv \Gamma_s + r\Gamma_\mathrm{ISC} \).
The relationship between the effective initialization rate \( \Gamma_{\mathrm{eff}} \) and laser power \( P \) is described by the saturation curve \( \Gamma_{\mathrm{eff}}(P) = (\alpha P + \Gamma_s^*)^{-1} \). For a given laser power, the observed photoluminescence as a function of time \( t \) is then modeled by
$
\mathrm{PL}(t;P) = \mathrm{PL}_\mathrm{ss}(P) (1 - e^{-\Gamma_{\mathrm{eff}}(P)t}),
$
where \( \mathrm{PL}_\mathrm{ss}(P) \) represents the steady-state photoluminescence at laser power \( P \).
Considering the Gaussian intensity distribution of the laser beam, the photoluminescence is calculated by taking the weighted average of \( \mathrm{PL}(t;P) \). In our fitting process, we treat \( \alpha \) and \( \Gamma_s^* \) as independent variables while keeping other transition rates fixed. 
\par
\begin{figure}[h]
    \centering
    \includegraphics[width=.8\linewidth]{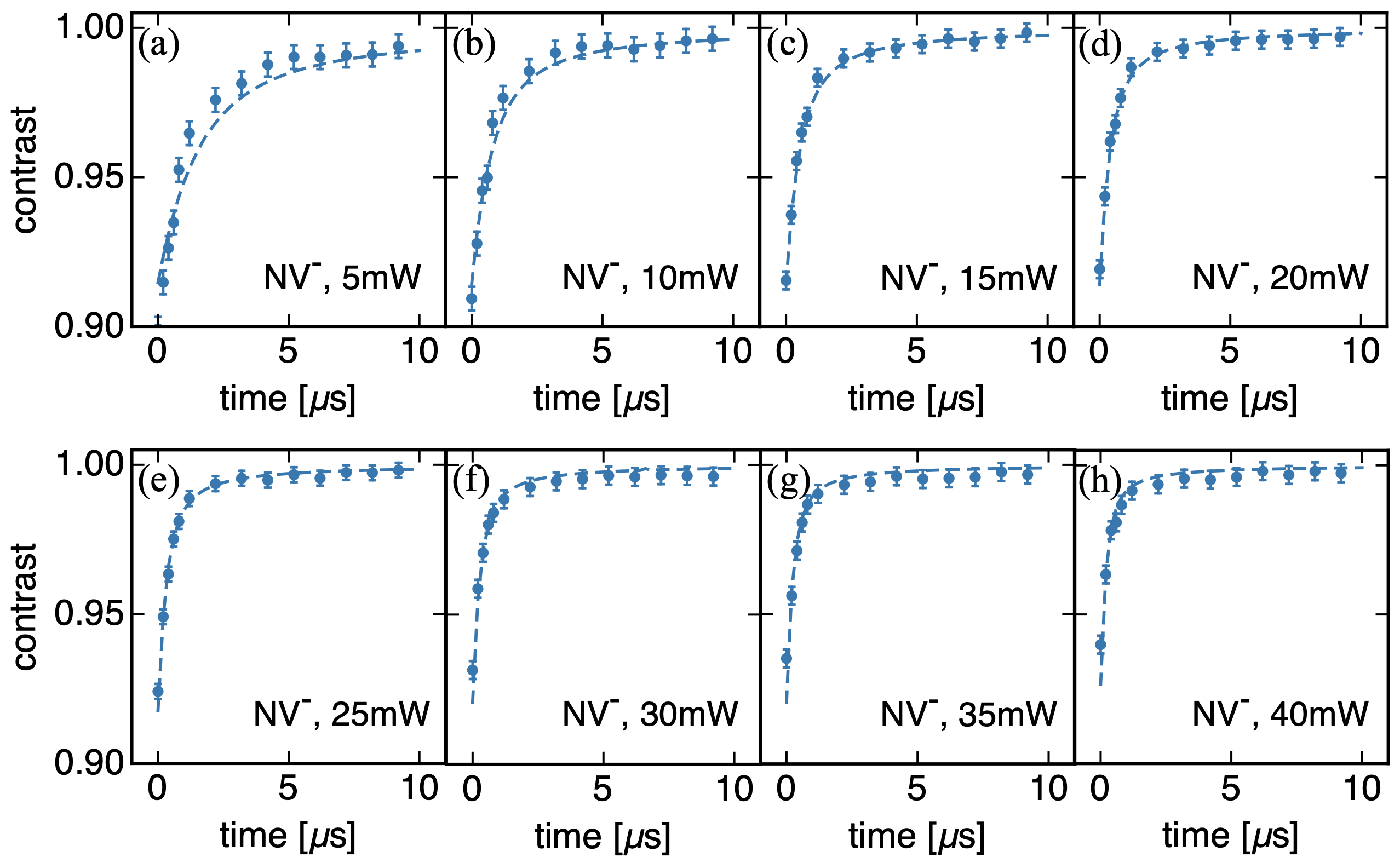}
    \caption{(a)-(h) $P_0$ signals of NV centers measured at different optical powers.}
    \label{fig:NV_r}
\end{figure}
Fig.~\ref{fig:NV_r} shows a set of measured $P_0$ signals at different laser powers, along with the corrsponding fitting results. As the laser power increases, the effective initialization rate \( \Gamma_{\mathrm{eff}} \) also increases. By simultaneously fitting these power-dependent photoluminescence data, we determine the optimal value of \( \Gamma_s^* \). These measurements were repeated across five different locations on the sample, yielding \( \Gamma_s^* = 7.8 \pm 2.6 \, \mathrm{MHz} \) and \( \alpha = 0.24 \pm 0.04 \, \mathrm{MHz}/\mathrm{mW} \).
Using the values of $\Gamma_s$ and $\Gamma_\mathrm{ISC}$ from [Table \ref{tab:NV_rate}], and the relationship \( \Gamma_s^* \equiv \Gamma_s + r\Gamma_\mathrm{ISC} \), we find $ r = 0.13 \pm 0.05$. This result is consistent with previous measurements presented in [Table \ref{tab:NV_rate}].

\begin{figure}[h]
    \centering
    \includegraphics[width=0.6\linewidth]{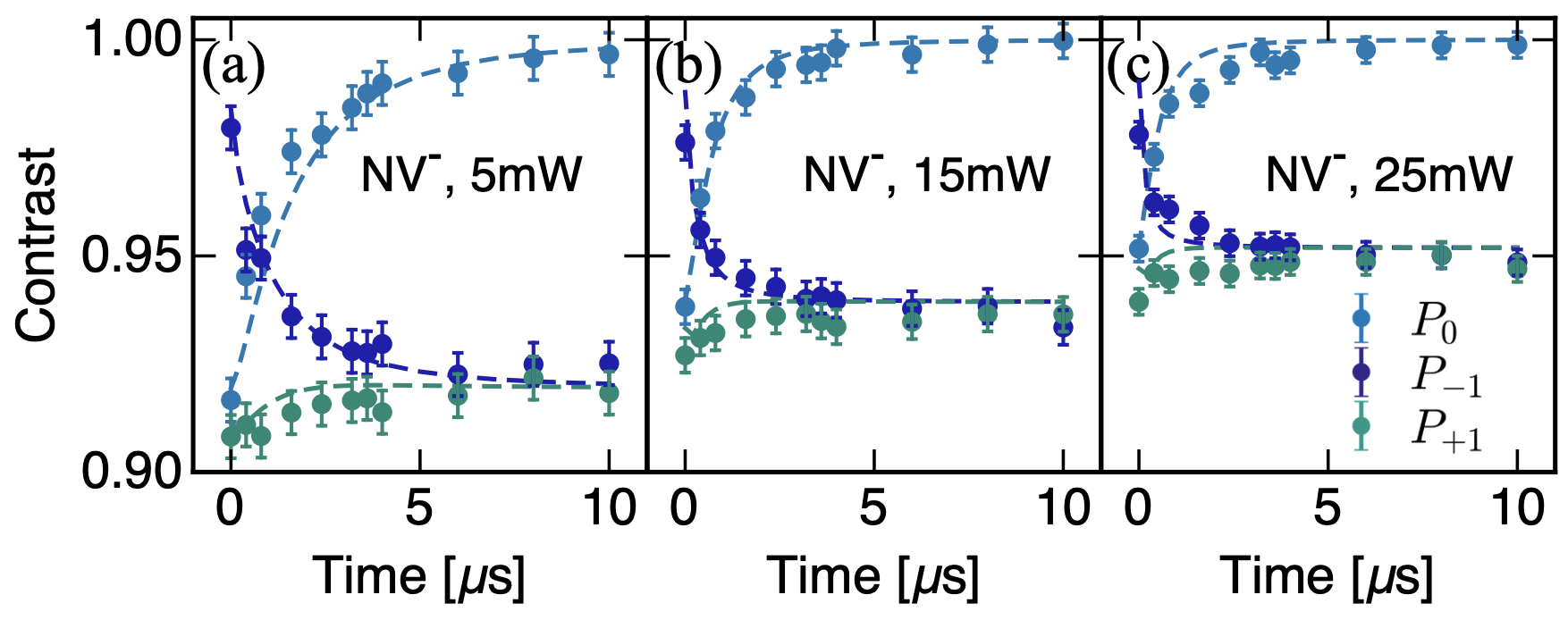}
    \caption{Spin-resolved polarization dynamics measurements for NV centers at different optical powers. The dashed lines represent the results obtained from the semiclassical model using transition rates from Table~\ref{tab:NV_rate}.}
    \label{fig:NV_BD}
\end{figure}

We also compare the full spin-resolved dynamics measurements, including $P_{\pm 1}$ signals, with theoretical predictions using the transition rates from Table~\ref{tab:NV_rate} [Fig.~\ref{fig:NV_BD}]. There are two notable differences compared to the fitting procedure used for the \vbm ~experiment. 
First, the fitted parameters are limited to the optical pumping rate \( \alpha \), the \(\pi\)-pulse fidelity \( f \), and the branching ratio $k$. Second, our measurements are averaged over a longer duration, specifically 500 ns, at the end of each pulse sequence. This is necessary in order to read out the ground state population for each spin state, given that the effective polarization rate for the NV center is lower than that of \vbm.
The fitting results are presented in Fig.~\ref{fig:NV_BD}, showing good agreement with the experimental data. 

\bibliography{supplement}